\documentclass[times,5p,twocolumn,sort&compress]{elsarticle}
\usepackage{graphicx}
\usepackage{amssymb}
\usepackage{amsmath}
\usepackage{amssymb}
\usepackage{url}
\journal{Journal of Chemical Physics; accepted for publication}
\begin{document}
\begin{frontmatter}
\title{On the Quantum Theory of Molecules}

\author[bts]{Brian T. Sutcliffe\corref{cor1}}
\ead{bsutclif@ulb.ac.be}

\author[rgw]{R. Guy Woolley}

\cortext[cor1]{Corresponding author}

\address[bts]{Service de Chimie quantique et Photophysique,  Universit\'{e} Libre de Bruxelles, B-1050 Bruxelles, Belgium}

\address[rgw]{School of Science and Technology, Nottingham Trent University, Nottingham NG11 8NS, U.K.}

\begin{abstract}
Transition state theory was introduced in the 1930s to account for
chemical reactions. Central to this theory is the idea of a potential
energy surface (PES). It was assumed that such a surface could be 
constructed using eigensolutions of the Schr\"{o}dinger equation for the molecular (Coulomb) Hamiltonian but at that time such calculations were not posssible. Nowadays quantum mechanical \emph{ab-initio} electronic structure calculations are routine and from their results PESs can be constructed which are believed to approximate those assumed derivable from the
eigensolutions. It is argued here that this belief is unfounded.
It is suggested that the potential energy surface construction is more appropriately regarded as a legitimate and effective modification of quantum mechanics for chemical purposes.
\end{abstract}

\end{frontmatter}

\def\mat#1{{\bf #1}} \def\gmat#1{\mbox{\boldmath$#1$}}
\newcommand{\be}{\begin{equation}}

\newcommand{\ee}{\end{equation}}

\newcommand{\rn}{{\rm n}}

\newcommand{\re}{{\rm e}}
\section{Introduction}
The principal aim of much of contemporary quantum chemical calculation
is to calculate a potential energy surface from solutions of the
Schr\"{o}dinger equation for the clamped-nuclei electronic Hamiltonian
which gives the electronic energy at a choice of nuclear positions. These electronic energies are then added to the clamped-nuclei classical Coulomb
repulsion energy and the resulting ``total'' energies fitted to a
functional form to provide a surface which, for a system of $A>2$
nuclei, is of dimension $3A-6$. Nuclear motion is treated as occurring
on this surface. 

The idea of a potential energy surface in such a role
began with the work of Eyring and the almost contemporary work of
Polanyi in the 1930s. Its basis is described in Chapter 16
of the textbook by Eyring, Walter and Kimball, published in 1944
\cite{EWK:44}. When describing the reaction between the hydrogen molecule and a hydrogen atom, the authors say:
\begin{quotation}
The properties of this system are completely determined by the
Schr\"{o}dinger equation governing it, that is by its
eigenfunctions. The system itself may be represented by a point in
four dimensional space: three dimensions are required to express the
relative positions of the nuclei, one additional dimension is required
to specify the energy. (We assume here that the motion of the
electrons is so rapid that the electrons form a static field for the
slower nuclear motions.) The three internuclear distances... can
conveniently be used to specify the configuration of the system.
\end{quotation}
\noindent They then give the name ``potential energy surface'' to the presumed calculated entity and go on to explain how transition probabilities for the hydrogen exchange reaction can be calculated using its
characteristic shape. They take it as obvious that the nuclei can be treated as classical particles whose positions can be fixed and do not mention the work of Born and Oppenheimer which, 17 years before the publication of their
book, had argued that such a treatment of the nuclei could be
justified if the potential energy surface had a unique minimum and
that the nuclear motion involved only small departures from
equilibrium \cite{BO:27,fn1}.

In the textbook by Pauling and Wilson \cite{PW:35} the work of Born
and Oppenheimer is quoted as justifying their use of a wavefunction
which is a simple product of an electronic and a nuclear part in order
to describe the vibration and rotation of molecules. They then
introduce the potential energy function, which they call the
``electronic energy function'', for a diatomic system. In their
description of a chemical reaction involving three atoms they attribute
the idea of fixing the nuclei to the work of London who used such an
approach in his 1928 paper \cite{L:28}. It is interesting to note that
London actually called this approach ``adiabatic'' saying that he
assumed that as the nuclei moved they acted as adiabatic parameters in
the electronic wavefunction.

Although treating the nuclei as classical particles that may be fixed
in space to generate a potential energy surface is often now referred
to as ``making the Born-Oppenheimer approximation'' it is something
of a mis-attribution as can be seen from this discussion. This was
indeed recognized by Born and in 1951 he published a paper \cite{B:51}
in which he presented a more general account of the separation of
electronic and nuclear motion than that originally offered. It is to
this later work by Born that the more general idea of a potential
energy surface is regarded as owing its justification and it is that
work that will be considered here. 

We aim to show that the potential energy surface does not arise naturally from the solution of the Schr\"{o}dinger equation for the molecular Coulomb Hamiltonian; rather its appearance requires the additional assumption that the nuclei can at first be treated as classical distinguishable particles and only later (after the potential energy surface has materialized) as quantum particles. In our view this assumption, although often very successful in practice, is \emph{ad hoc}.  

The outline of the paper is as follows. In the next section we review the main features of the standard Born-Oppenheimer and Born adiabatic treatments following Born and Huang's well-known book \cite{BH:54}. The key idea is that the nuclear kinetic energy contribution can be treated as a small perturbation of the electronic energy; the small parameter ($\kappa$) in the formalism is obtained from the ratio of the electronic mass ($m$) to the nuclear mass ($M_o$): $m/M_o~=~\kappa^4$. The argument leads to the expression of the molecular Hamiltonian as the sum of the ``clamped-nuclei'' electronic Hamiltonian (independent of $\kappa$) and the nuclear kinetic energy operator ($\propto~\kappa^4$), equation (\ref{Hpert}).

In \S \ref{modth} we attempt a careful reformulation of the conventional Born-Oppenheimer argument drawing on results from the modern mathematical literature. The calculation is essentially concerned with the \emph{internal motion} of the electrons and nuclei so we require the part of the molecular Hamiltonian that remains after the center-of-mass contribution has been removed. We show that it is possible to express the internal motion Hamiltonian in a form analogous to equation (\ref{Hpert}); however the electronic part, independent of $\kappa$, is \emph{not} the clamped-nuclei Hamiltonian. Instead, the exact electronic Hamiltonian can be expressed as a direct integral of clamped-nuclei Hamiltonians and necessarily has a \emph{purely continuous spectrum of energy levels}; there are \emph{no} potential energy surfaces. This continuum has nothing to do with the molecular centre-of-mass, by construction.  The paper concludes (\S \ref{Disc}) with a discussion of our finding.

\section{The Born-Oppenheimer approximation}
\label{tradBO}
The original Born and Oppenheimer approximation \cite{BO:27,fn1} is summarized in the famous book by Born and Huang, and
the later Born adiabatic method \cite{B:51} is given in an appendix to
that book \cite{BH:54}. Born and Huang use the same notation for both
formulations and it is convenient to follow initially their
presentation; the following is a short account focusing on the main ideas. They work in a position representation and for simplicity suppress all individual particle labels. Let us consider a system of electrons and nuclei and denote the properties of the former by lower-case letters (mass $m$, coordinates $x$, momenta $p$) and of the latter by capital letters (mass $M$, coordinates $X$, momenta
$P$). The kinetic energy of the nuclei is the operator 
\begin{equation}
T_N~=~\sum \frac{1}{2M}~P^2~=~-~\sum
\frac{\hbar^2}{2M}\left(\frac{\partial^2}{\partial X^2}\right) 
\label{nuke}
\end{equation}
and that of the electrons
\begin{equation}
T_e~=~\sum \frac{1}{2m}~p^2~=~-~\sum
\frac{\hbar^2}{2m}\left(\frac{\partial^2}{\partial x^2}\right). 
\label{eke}
\end{equation}
The total Coulomb energy of the electrons will be represented by
$U(x,X)$. We further introduce the abbreviation 
\begin{equation}
T_e~+~U~=~H_o\left(x,\frac{\partial}{\partial x}, X\right).
\label{hcn}
\end{equation}
The full Hamiltonian for the molecule is then 
\begin{equation}
H~=~T_e~+~U~+T_N~=~H_o~+~T_N.
\label{hfull}
\end{equation}

The fundamental idea of Born and Oppenheimer is that the low-lying
excitation spectrum of a typical molecule can be calculated by
regarding the nuclear kinetic energy $T_N$ as a small perturbation of
the Hamiltonian $H_o$. The physical basis
for the idea is the large disparity between the mass of the electron
and all nuclear masses. The expansion parameter must clearly be some
power of $m/M_o$, where $M_o$ can be taken as any one of the nuclear
masses or their mean. They found that the correct choice is 
\begin{displaymath}
\kappa~=~\left(\frac{m}{M_o}\right)^{\frac{1}{4}}
\end{displaymath}
and therefore
\begin{equation}
T_N~=~\kappa^4~H_1\left(\frac{\partial}{\partial X}\right)~=~\sum
\left(\frac{M_o}{M}\right)~
\frac{\hbar^2}{2m}~\left(\frac{\partial^2}{\partial X^2}\right).
\label{nkekap}
\end{equation}
Thus the total Hamiltonian may be put in the form
\begin{equation}
H~=~H_o~+~\kappa^4~H_1
\label{Hpert}
\end{equation}
with Schr\"{o}dinger equation
\begin{equation}
\big(H~-~E\big)\psi(x,X)~=~0.
\label{scheq}
\end{equation}

In the original paper Born and Oppenheimer say at this
point in their argument that \cite{fn1}:
\begin{quotation}
If one sets $\kappa=0$ ... one obtains a differential equation in the
${x}$ alone, the ${X}$ appearing as parameters:
\[~\left[H_o\left(x,\frac{\partial}{\partial x}, X\right)-W\right]\psi=0.\]
This represents the electronic motion for stationary nuclei.
\end{quotation}
and it is perhaps to this statement that the idea of an electronic 
Hamiltonian with fixed nuclei as arising by letting the nuclear masses
increase without limit, can be traced. In modern parlance ${H}_o$ is
customarily referred to as the ``clamped-nuclei Hamiltonian''.

Consider the unperturbed electronic Hamiltonian $H_o(x,X_f)$ at a
fixed nuclear configuration $X_f$ that corresponds to some molecular
structure. The Schr\"{o}dinger equation for $H_o$ is 
\begin{equation}
\bigg(H_o(x,X_f)~-~E^{o}(X_f)_m\bigg)\varphi(x,X_f)_m~=~0.
\label{scheqcn}
\end{equation}
 This Hamiltonian's natural domain, ${\cal{D}}_o$, is the
set of square integrable electronic wavefunctions \{$\varphi_m$\} with
square  integrable first and second derivatives; ${\cal{D}}_o$ is
independent of $X_f$. We may suppose the \{$\varphi_m$\} are
orthonormalized independently of $X_f$
\begin{displaymath}
\int dx ~\varphi(x, X_f)_{n}^{*}~\varphi(x, X_f)_m~=~\delta_{nm}.
\end{displaymath}
In the absence of degeneracies (``curve-crossing'') they may be chosen
to be real; otherwise there is a phase factor to be considered.  For every $X_f$, $H_o$ is self-adjoint on the electronic Hilbert space
${\cal{H}}(X_f)$, and therefore the set of states \{$\varphi(x, X_f)_m$\} form a complete set for the electronic Hilbert space indexed by $X_f$.

\begin{figure}[htbp]
\centering \includegraphics[width=3in]{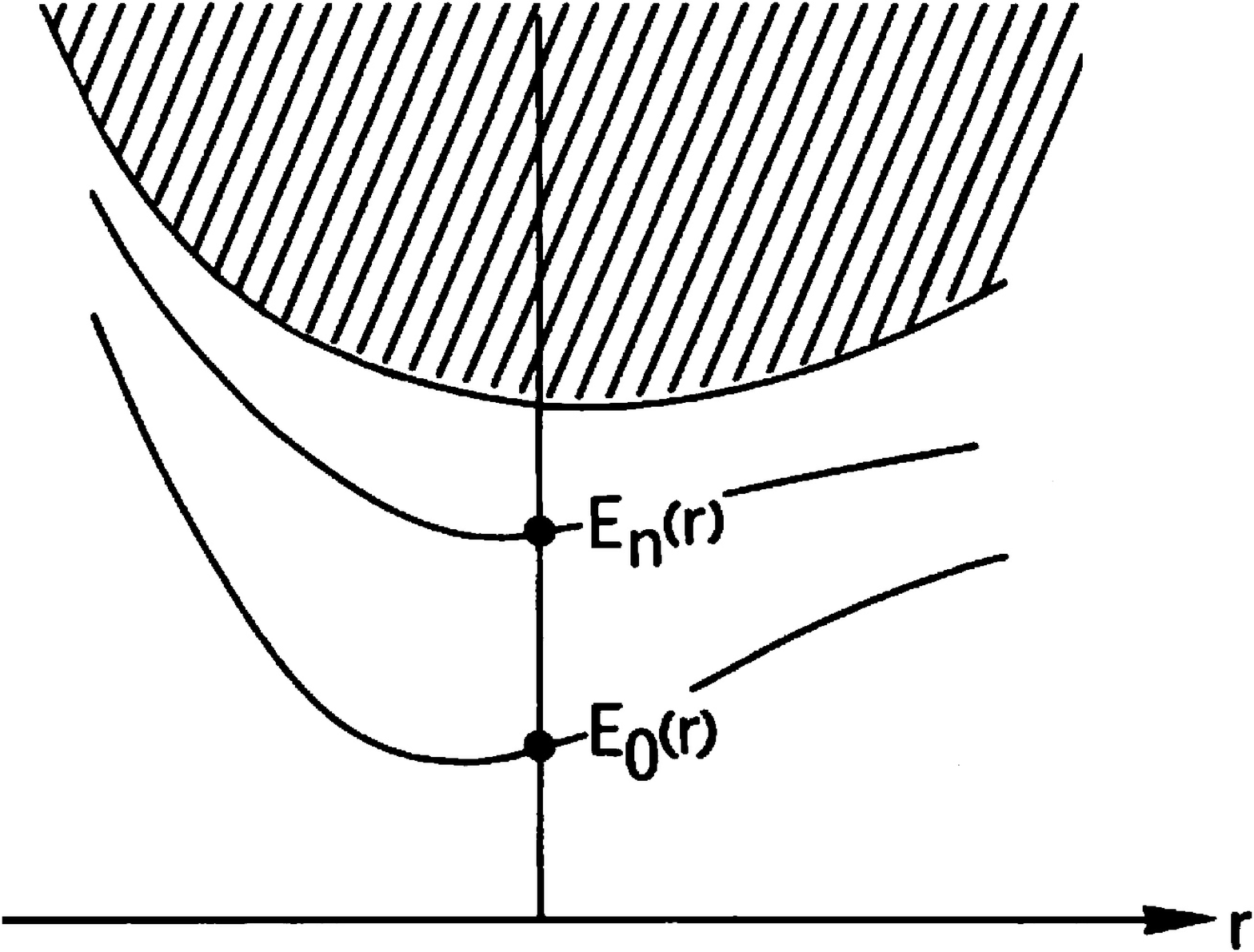}
\caption{The spectrum $\sigma(r)$ for a diatomic molecule
\cite{CS:80}.} \label{PEcurve}
\end{figure}

The clamped-nuclei Hamiltonian can be analysed with the HVZ theorem which shows that it has both discrete and continuous parts to its spectrum \cite{CS:80,RS:78,ZH:60,Uch:67,Th:81},
\begin{equation}~
  \sigma(X_f)\equiv~\sigma(H_o(x,X_f))=\bigg[E^{o}(X_f)_0, \ldots
    E^{o}(X_f)_m\bigg)\bigcup \bigg[\Lambda(X_f),\infty\bigg) 
\label{speccn}
\end{equation}
where the \{$E^{o}(X_f)_k$\} are isolated eigenvalues of finite
multiplicities. $\Lambda(X_f)$ is the bottom of the essential spectrum
marking the lowest continuum threshold. In the case of a diatomic
molecule the electronic eigenvalues depend only on the internuclear
separation $r$, and have the form of the familiar potential curves
shown in Fig.\ref{PEcurve}. 
For the general polyatomic molecule, the discrete eigenvalues are molecular potential energy surfaces.

Born and Oppenheimer used the set \{$\varphi_m$\} to calculate
approximate eigenvalues of the full molecular Hamiltonian $H$ on the
assumption that the nuclear 
motion is confined to a small vicinity of a special (equilibrium)
configuration $X_f^0$. Their great success was in establishing that
the energy levels of the low-lying states typical of small polyatomic
molecules obtained from molecular spectroscopy could be written as an
expansion in powers of $\kappa^2$ 
\begin{equation}
E_{nvJ}~\approx~
V_n^{(0)}~+~\kappa^2~E_{nv}^{(2)}~+~\kappa^4~E_{nvJ}^{(4)}~+~\ldots
\label{enlevel}
\end{equation}
where $V_n^{(0)}$ is the minimum value of the electronic energy which
characterises the molecule at rest, $E_{nv}^{(2)}$ is the energy of
the nuclear vibrations, and $E_{nvJ}^{(4)}$ contains the rotational
energy \cite{BO:27}. The corresponding approximate wavefunctions are
simple products of an electronic function $\varphi_m$ and a nuclear
wavefunction; this is known as \emph{the adiabatic approximation}. In
the original perturbation formulation the simple product form is valid
through $\kappa^4$, but not for higher order terms.

About 25 years later Born observed that the results of molecular
spectroscopy suggest that the adiabatic approximation has a wider
application than predicted by the original theory, and he proposed an
alternative formulation \cite{B:51,BH:54}. It is assumed that
the functions $E^o(X)_m$ and $\varphi(x,X)_m$ arising from equation
(\ref{scheqcn}) which represent the energy and wavefunction of the
electrons in the state $m$ for a fixed nuclear configuration $X$, are
known.  Born proposed to solve the wave equation (\ref{scheq}) by an
expansion
\begin{equation}
\psi(x,X)~=~\sum_m \Phi(X)_m~\varphi(x,X)_m
\label{wfnexpan}
\end{equation}
with coefficients \{$\Phi(X)_m$\} that play the role of nuclear
wavefunctions. Substituting this expansion into the full
Schr\"{o}dinger equation (\ref{scheq}), multiplying the result by
$\varphi(x,X)_n^{*}$ and integrating over the electronic coordinates
$x$ leads to a system of coupled equations for the nuclear functions
\{$\Phi$\},
\begin{equation}
\big(T_N~+~E^{o}(X)_n~-~E\big)\Phi(X)_n~+~\sum_{nn'}C(X,P)_{nn'}\Phi(X)_{n'}~=~0
\label{coupl}
\end{equation}
where the coupling coefficients \{$C(X,P)_{nn'}$\} have a well-known
form which we need not record here \cite{BH:54}. In this formulation the adiabatic approximation consists of retaining only the diagonal terms in the coupling matrix ${\bf C}(X,P)$, for then
\begin{equation}
\psi(x,X)~\approx~\psi(x,X)_{n}^{\mbox{AD}}~=~\varphi(x,X)_n~\Phi(X)_n.
\end{equation}

An obvious defect in this presentation, recognized by Born and Huang  \cite{BH:54}, is that the kinetic energy of the overall center-of-mass is retained in the Schr\"{o}dinger equation (\ref{scheq}); this is easily corrected, either by the explicit separation of the center-of-mass kinetic energy operator (see \S \ref{modth}), or implicitly, as in the computational scheme proposed by Handy and co-workers \cite {HYS:86,HL:96,K:97}. For many years now these equations have been regarded in the theoretical molecular spectroscopy/quantum chemistry literature as defining the
``Born-Oppenheimer approximation'', the original perturbation method
being relegated to the status of historical curiosity. Commonly they
are said to provide an exact (in principle) solution \cite{AM:60,MM:67,TOM:71,HJM:87,LT:10}  for the
stationary states of the molecular Schr\"{o}dinger equation
(\ref{scheq}), it being recognized that in practice drastic truncation
of the infinite set of coupled equations (\ref{coupl}) is required.
 In the next section we make a critical evaluation of this conventional account.

\section{The molecular Schr\"{o}dinger equation}
\label{modth}
We now start again and reconsider the Hamiltonian for a collection of
electrons and nuclei, assuming their interactions are restricted to
the usual Coulombic form. The key idea in \S \ref{tradBO} is the decomposition of the molecular Hamiltonian (\ref{hfull}) into a part containing all contributions of the nuclear momenta, and a remainder. This must be done in conjunction with a proper treatment of the center-of-mass motion. These two ideas guide the following discussion.

Let the position variables for the
particles in a laboratory fixed frame be designated as \{${\bf
  x}_i$\}. When it is necessary to distinguish between electrons and
nuclei, the position variables may be split up into two sets, one set
consisting of $N$ coordinates, ${\mat x}^{\rm e}_i$, describing the
electrons with charge $-e$ and mass $m$, and the other set of $A$
coordinates, ${\bf x}^{\rm n}_g$, describing the nuclei with charges
$+Z_ge$ and masses $m_g$, $g=1,\dots A$; then $n=N+A$. We define
corresponding canonically conjugate momentum variables \{${\bf p}_i$\}
for the electrons and nuclei. With this notation, and after the usual
canonical quantization, the Hamiltonian operator for a system of $N$
electrons and $A$ atomic nuclei with Coulombic interactions may be
written as
\begin{eqnarray}
\mathsf{H}&=&\sum_{g}^{A}\frac{\mathsf{p}_g^2}{2m_g}+\frac{e^2}{4\pi\epsilon_0}\sum_{g<h}^A\frac{Z_gZ_h}{\mathsf{r}_{gh}}
+\frac{e^2}{4\pi\epsilon_0}\sum_{i<j}^N\frac{1}{\mathsf{r}_{ij}}\nonumber\\ 
&+&\sum_i^N\left(\frac{\mathsf{p}_i^2}{2m}-\frac{e^2}{4\pi\epsilon_0}\sum_{g}^A\frac{Z_g}{\mathsf{r}_{ig}}\right).
\label{hfullop}
\end{eqnarray}

It is easily seen that the Coulomb interaction is translation
invariant. Thus the total momentum operator, which in the present
notation  is
\begin{displaymath}
\mathsf{\bf P} ~=~\sum_i^n \mathsf{\bf p}_i,
\end{displaymath}
commutes with $\mathsf{H}$. Physically the center-of-mass of the whole system, with
position operator
\begin{displaymath}
\mathsf{\bf R}=\frac{1}{M_T}\sum_i^nm_i \mathsf{\bf x}_i,~~M_T=\sum_i^nm_i \
\end{displaymath}
behaves like a free particle. It is desirable then to introduce
$\mathsf{\bf R}$ and its conjugate $\mathsf{\bf P}_{{\bf R}}$, together with
appropriate internal coordinates, into $\mathsf{H}$ to make explicit the
separation of the center-of-mass and the internal dynamics. Formally
$\mathsf{H}$ may be written as a direct integral \cite{fn2}
\begin{equation}
\mathsf{H}~=~\int_{\mbox{\bf R}^3}^{\oplus}H(P)~dP
\label{dirint}
\end{equation}
where \cite{LMS:07}
\begin{equation}
H(P)=\frac{{P}^2}{2M_T} +\mathsf{H}'
\label{sep}
\end{equation}
is the Hamiltonian at fixed total momentum $P$. This representation shows directly that $\mathsf{H}$ has purely continuous spectrum. The
internal Hamiltonian $\mathsf{H}'$ is independent of the center-of-mass
variables and acts on $L^2({\bf R}^{3(n-1)})$; it is explicitly
translation invariant.

There are infinitely many possible choices of internal coordinates
that are unitarily equivalent, so that the form of $\mathsf{H}'$ is not
determined uniquely, but whatever coordinates are chosen the essential
point is that $\mathsf{H}'$ is the \emph{same} operator specified by the
decomposition (\ref{sep}). A simple procedure to make the internal
Hamiltonian explicit is to refer the particle coordinates to a point
moving with the system, for example, the center-of-mass itself, the
center-of-nuclear-mass or one of the moving particles
\cite{POL:88}.

As a result of the transformation to internal variables the kinetic
energy operators are no longer diagonal in the particle indices and
certain choices of the moving point, such as the choice of a single
nucleus, result in an operator in which the nuclear and electronic
indices are mixed. However the choice of the center-of-nuclear-mass as
the point of origin avoids this mixing; the explicit equations for
this choice were given in Sutcliffe and Woolley \cite{SW:05a}. 
There are $A-1$ translationally invariant coordinates ${\bf t_{i}^{\rm n}}$ expressed entirely in terms of the original ${\bf x}_{g}^{\rm n}$ that may be
associated with the nuclei, and there are N translationally invariant
coordinates for the electrons ${\bf t}_{i}^{\rm e}$ which are simply
the original electronic coordinates ${\bf x}_{i}^{\rm e}$ referred to
the center-of-nuclear-mass. There are corresponding canonically
conjugate internal momentum operators. The classical total kinetic energy still separates in the form
\begin{displaymath}
\mathsf{T}_0~\equiv~T_N~+~T_e~=~\mathsf{T}_{CM}~+~\mathsf{T}_{Nu}~+~\mathsf{T}_{el}
\end{displaymath}
and the same is true after quantization. $\mathsf{T}_{CM}$ is the kinetic
energy for the center-of-mass and, for example, $\mathsf{T}_{el}$ only involves electronic variables 
\begin{equation}
\mathsf{T}_{el}~=~\frac{1}{2\mu}\sum_{i=1}^N(\pi^{\rm e}_i)^2~+~
\frac{1}{2M}\sum_{ij=1}^N\!\hbox{\raisebox{5pt}{${}^\prime$}} 
(\gmat{\pi}^{\rm e}_i) \cdot (\gmat{\pi}^{\rm e}_j)
\label{intke}
\end{equation}
with 
\begin{displaymath}
M~=~\sum_{g=1}^A m_g,~~~~~\frac{1}{\mu}~=~\frac{1}{m}~+~\frac{1}{M}.
\end{displaymath}
The nuclear kinetic energy term is like the second term in
(\ref{intke}) expressed in terms of internal nuclear momentum
variables \{$\gmat{\Pi}_g^{\rm n}$\} with a mass factor in the
denominator composed from the nuclear masses.

After this unitary transformation the original Coulomb Hamiltonian operator for the molecule can be rewritten in the form 
\begin{equation}
\mathsf{H}(\mathsf{\bf x}^{\rm e},\mathsf{\bf p}^{\rm e},\mathsf{\bf x}^{\rm
  n},\mathsf{\bf p}^{\rm
  n})~=~\mathsf{T}_{CM}~+~\mathsf{T}_{Nu}~+~\mathsf{H}^{\mbox{elec}} 
\label{sepeqn}
\end{equation}
where $\mathsf{H}^{\mbox{elec}}$ is composed of (\ref{intke}) together
with all the Coulomb interaction operators expressed in terms of the
\{${\bf t}^{\rm e},{\bf t}^{\rm n}$\} position operators.

The spectrum of the Coulomb operator $\mathsf{H}'$ formed by
dropping $~\mathsf{T}_{CM}~$ from (\ref{sepeqn}) was considered
by Kato \cite{K:51} who showed in Lemma 4 of his paper that for a Coulomb potential $\mathsf{V}$ and for any function $f$ in the domain
$\mathcal{D}_0$ of the full kinetic energy operator $\mathsf{T}_0$,
the domain, $\mathcal{D}_V$, of the internal Hamiltonian $\mathsf{H}'$ 
 contains $\mathcal{D}_0$ and there are two constants
$a,~b$ such that 
\[ ||\mathsf{V}f|| \leq a||\mathsf{T}_0 f|| +b||f|| \]
where $a$ can be taken as small as is liked.  This result is often
summarised by saying that the Coulomb potential is small compared to
the kinetic energy. Given this result he proved in Lemma 5 (the
Kato-Rellich theorem) that the usual operator is indeed, for all
practical purposes, self-adjoint  and so is guaranteed a complete set of eigenfunctions, and is bounded from below. 

In the present context the important point to note is that the Coulomb
term is small only in comparison with the kinetic energy term
involving the same set of variables. So the absence of one or more 
kinetic energy terms from the Hamiltonian may mean that the Coulomb
potential term cannot be treated as small. Such a defective
Hamiltonian will no longer be self-adjoint in the way demonstrated by
Kato.

With our choice of coordinates the translationally invariant
Hamiltonian may be written  as the sum of the last two terms in
(\ref{sepeqn}) and the electronic Hamiltonian now becomes
\begin{eqnarray}\mathsf{H}^{\mbox{elec}} &=&
-\frac{\hbar^2}{2m}\sum_{i=1}^{N}
{\nabla}^2({\mat t}^{\rm e}_i)
-\frac{\hbar^2}{2 M}\sum_{i, j=1}^{N}
\vec{\nabla}(\mat{t}^{\rm e}_i)\cdot\vec{\nabla}(\mat{t}^{\rm e}_j)\nonumber\\  &-&\frac{e^2}{4\pi{\epsilon}_0}\sum_{i=1}^A
\sum_{j=1}^N\frac{Z_i}{|{\mat t}^{\rm e}_j - {\mat x}^{\rm n}_i|}\nonumber\\ 
&+& \frac{e^2}
{8\pi{\epsilon}_0}\sum_{i,j=1}^N\!\hbox{\raisebox{5pt}{${}^\prime$}}
\frac{1}{|{\mat t}^{\rm e}_i - {\mat t}^{\rm e}_j|} + \sum_{i,j=1}^A\!
\hbox{\raisebox{5pt}{${}^\prime$}} \frac{Z_iZ_j}{|\mat x^{\rm n}_i -
  \mat x^{\rm n}_j|}
\label{hemol}
\end{eqnarray}
where it is understood that the $\mat{x}^{\rn}_i$ are to be realised
by a suitable linear combination of the $\mat{t}^{\rn}_i$.
The electronic Hamiltonian is properly translationally invariant and, were the nuclear masses to increase without limit, the second tem in (\ref{hemol}) would vanish and the electronic Hamiltonian would have the same form as that usually invoked in molecular electronic stucture calculations. If the nuclear positions were chosen directly as a translationally invariant set, it
would be those values which would appear in the place of the nuclear
variables.
The nuclear part involves only kinetic energy operators and has
the form:
\begin{equation}
\mathsf{T}_{Nu} = -\frac{\hbar^2}{2}\sum_{i,j=1}^{A-1}\frac{1}{{\mu}_{ij}}\vec{\nabla}({\mat t}^{\rm n}_i).\vec{\nabla}({\mat t}^{\rm n}_j)
\label{hnt}
\end{equation}
with the inverse mass matrix $\gmat{\mu}$ given in terms of the
coefficients that define the translationally invariant coordinates and
the nuclear masses.

$\mathsf{H}^{\mbox{elec}}$ has the same invariance under the
rotation-reflection group O(3) as does the
full translationally invariant Hamiltonian $\mathsf{H}'$ and, in the case
of the molecule containing some identical nuclei, it has somewhat
extended invariance under nuclear permutations, since the nuclear
masses appear only in symmetrical sums. In a position representation
the Schr\"{o}dinger equation for $\mathsf{H}^{\mbox{elec}}$ may be
written as 
\begin{equation}
\mathsf{H}^{\mbox{elec}}~\Psi({\bf t}^{\rm e},{\bf t}^{\rm n})_{\bf
  m}~=~{\cal{E}}_{\bf m}~\Psi({\bf t}^{\rm e},{\bf t}^{\rm n})_{\bf m}  
\label{eigeq}
\end{equation}
where ${\bf m}$ is used to denote a set of quantum numbers $(JMprk)$:
$J$ and $M$ for the angular momentum state: $p$ specifying the parity
of the state: $r$ specifying the permutationally allowed irreps within
the group(s) of identical particles and $k$ to specify a particular
energy value \cite{SW:05a}. 

Before discussing (\ref{eigeq}) for the general case, let us consider a specific example for which high-quality computational results exist. It might be hoped, in the light of the claim in the original paper by Born and Oppenheimer quoted in \S \ref{tradBO}, that the solutions of (\ref{eigeq}) would actually be those that would have been obtained from an exactly solved problem by letting the nuclear masses increase without limit. Although there are no exactly solved molecular problems, the work of Frolov
\cite{FROL:99} provides extremely accurate numerical solutions for a
problem with two nuclei and a single electron. Frolov investigated what happens when the masses of one and then two of the nuclei increase without limit in his calculations. To appreciate his results, consider a system with two nuclei; the natural nuclear coordinate is the internuclear distance which will be denoted here simply as $\mat{t}$.  When needed to express the electron-nuclei attraction terms, $\mat{x}^{\rm n}_i$ is simply of the form 
$\alpha_i\mat{t}$ where $\alpha_i$ is a signed ratio of the nuclear mass
to the total nuclear mass; in the case of a homonuclear system
$\alpha_i=\pm\frac{1}{2}$. 

The di-nuclear electronic Hamiltonian after the elimination of the center-of-mass contribution is 
\begin{eqnarray}
\!\!\!\!\mathsf{H}^{' \rm e}({\mat t}^{\rm e}) &=& -\frac{\hbar^2}{2m}\sum_{i=1}^{N} 
{\nabla}^2({\mat t}^{\rm e}_i) -\frac{\hbar^2}{2(m_1+m_2)}\sum_{i, j=1}^{N} 
\vec{\nabla}(\mat{t}^{\rm e}_i)\cdot\vec{\nabla}(\mat{t}^{\rm e}_j)\nonumber\\&-&\frac{e^2}{4\pi{\epsilon}_0}\sum_{j=1}^N \left(\frac{Z_1}{|{\mat t}^{\rm
e}_j+\alpha_1\mat{t}|}+\frac{Z_2}{|{\mat t}^{\rm e}_j+\alpha_2\mat{t}|}\right)\nonumber\\ &+&\frac{e^2}{8\pi{\epsilon}_0}\sum_{i,j=1}^N\!\hbox{\raisebox{5pt}{${}^\prime$}}\frac{1}{|{\mat t}^{\rm e}_i - {\mat t}^{\rm e}_j|}+\frac{Z_1Z_2}{R},~~R=|\mat{t}| 
\label{dineh}
\end{eqnarray}
while the nuclear kinetic energy part is:
\be
\mathsf{T}_{Nu}(\mat t)~=~-\frac{\hbar^2}{2}\left(\frac{1}{m_1}+\frac{1}{m_2}\right){\nabla}^2({\mat t}) \equiv-\frac{\hbar^2}{2\mu}{\nabla}^2({\mat t}). 
\label{dinke}
\ee
The full internal motion Hamiltonian for the three-particle system is then
\be
\mathsf{H}'({\mat t}^{\rm e}, \mat{t})~=~\mathsf{H}^{' \rm e}({\mat t}^{\rm e}) ~+~\mathsf{T}_{Nu}(\mat{t}).
\label{intH}
\ee

It is seen from (\ref{dinke}), that if only one nuclear mass increases
without limit then the kinetic energy term in the nuclear variable
remains in the full problem and so the Hamiltonian (\ref{intH}) remains

self-adjoint in the Kato sense. Frolov's calculations showed that when one mass increased without limit (the atomic case), any discrete spectrum persisted but when two masses were allowed to increase without limit (the molecular case), the Hamiltonian ceased to be well-defined and this failure led to what he called \emph{adiabatic divergence} in attempts  to compute discrete eigenstates of ({\ref{intH}). This divergence is discussed in some mathematical detail in the Appendix to Frolov\cite{FROL:99}. It does not arise from the choice of a translationally invariant form for the electronic Hamiltonian; rather it is due to the lack of any kinetic energy term to dominate the Coulomb potential. Thus it really is essential to characterize the spectrum of $\mathsf{H}^{\mbox{elec}}$ to see whether the traditional approach can be validated.

As before, the nuclear kinetic energy operator is proportional to $\kappa^4$, so after dropping the uninteresting center-of-mass kinetic energy term, (\ref{sepeqn}) is seen to be of the same form as (\ref{Hpert}). There is however a fundamental difference between (\ref{Hpert}) and (\ref{sepeqn}) which may be seen as follows; with the center-of-nuclear-mass chosen as the electronic origin $\mathsf{H}^{\mbox{elec}}$ is independent of the nuclear
momentum operators and so it commutes with the nuclear position
operators
\begin{displaymath}
[\mathsf{H}^{\mbox{elec}}, \mathsf{\bf t}^{\rm n}]~=~0.
\end{displaymath}
They may therefore be simultaneously diagonalized and we use this
property to characterize the Hilbert space ${\cal{H}}$ for
$\mathsf{H}^{\mbox{elec}}$. Let ${\bf b}$ be some eigenvalue of the ${\bf
  t}^{\rm n}$ corresponding to choices \{${\bf x}_g~=~{\bf a}_g, g = 1,\ldots A$\} in the laboratory-fixed frame; then the \{${\bf a}_g$\} describe a classical nuclear geometry. The set, $X$, of all ${\bf b}$ is ${\bf
  R}^{3(A-1)}$. We denote the Hamiltonian $\mathsf{H}^{\mbox{elec}}$
evaluated at the nuclear position eigenvalue ${\bf b}$ as
$\mathsf{H}({\bf b},{\bf t}^{\rm e})_o~=~\mathsf{H}_o$ for short; this
$\mathsf{H}_o$ is very like the usual clamped-nuclei Hamiltonian but it
is explicitly translationally invariant, and has an extra term, the
second term in (\ref{intke}), (or (\ref{dineh})), which is often called the Hughes-Eckart term, or the mass polarization term. The Schr\"{o}dinger equation for  $\mathsf{H}_o$ is of the same form as (\ref{scheqcn}), with eigenvalues $E^{o}({\bf  b})_k$ and corresponding eigenfunctions $\varphi({\bf t}^{\rm e},{\bf b})_k$, and with spectrum $\sigma({\bf b})$ analogous to (\ref{speccn}).

In general the symmetries of $\mathsf{H}({\bf b},{\bf t}^{\rm e})_o$
considered as a function of the electronic coordinates, ${\bf t}^{\rm
  e}$, are much lower than those of $\mathsf{H}^{\mbox{elec}}$, since they
are determined by the (point group) transformations that leave the
geometrical figure defined by the \{${\bf a}_g$\} invariant. In a chiral
structure there is no symmetry operation other than the identity so
even space-inversion symmetry (parity) is lost.  

At any choice of {\bf b} the eigenvalues of $\mathsf{H}_o$ will depend
only upon the shape of the geometrical figure formed by the \{${\bf a}_g$\}
and not at all upon its orientation. For other than diatomic
molecules, it is technically possible to produce a new coordinate
system from the {\bf b} with angular variables describing 
the orientation of the figure and 3A-6 internal variables describing its
geometry. However this cannot be done without angular momentum terms
arising in which electronic and nuclear momentum operators are
coupled. Thus the kinetic energy can no longer be expressed as a sum
of separate electronic and nuclear parts. There are also many possible
choices of angular and internal variables which span different
(though often overlapping) domains, so it is not possible to produce a
single account of the separation of electronic and nuclear motion,
applicable to all choices.

As in \S \ref{tradBO} $\mathsf{H}_o$ is self-adjoint on an electronic
Hilbert space $\cal{H}({\bf b})$, so we have a family of Hilbert
spaces \{$\cal{H}({\bf b})$\} which are parameterized by the nuclear
position vectors ${\bf b} \in X$ that are the ``eigenspaces'' of the
family of self-adjoint operators $\mathsf{H}_o$; from them we can
construct a big Hilbert space as a direct integral over all the ${\bf
  b}$ values 
\begin{equation}
{\cal{H}}~=~\int^{\oplus}_{X}{\cal{H}}({\bf b})~d{\bf b}
\label{bigH}
\end{equation}
and this is the Hilbert space for $\mathsf{H}^{\mbox{elec}}$ in
(\ref{sepeqn}). The internal molecular Hamiltonian $\mathsf{H}'$ in
(\ref{sep}) and the clamped-nuclei like operator $\mathsf{H}_o$ just
defined can be shown to be self-adjoint (on their respective Hilbert
spaces) by reference to the Kato-Rellich theorem \cite{RS:78} because
in both cases there are kinetic energy operators that dominate the
(singular) Coulomb interaction; they therefore have a complete set of
eigenfunctions. This argument cannot be made for
$\mathsf{H}^{\mbox{elec}}$ because it contains nuclear position operators
in some Coulombic terms but there are no corresponding nuclear kinetic
energy terms to dominate those Coulomb potentials. It therefore cannot be thought of as being self-adjoint on the full Hilbert space of electronic and nuclear variables. Thus even though
$\mathsf{H}^{\mbox{elec}}$ is evidently Hermitian symmetric, this is not
sufficient to guarantee a complete set of eigenfunctions because the operator is unbounded, and an expansion analogous to (\ref{wfnexpan}) seems quite problematic.

However that may be, equation (\ref{bigH}) leads directly to a
fundamental result; since $\mathsf{H}^{\mbox{elec}}$ commutes with all
the \{$\mathsf{{\bf t}}^{\rm n}$\}, it  has the direct integral
decomposition 
\begin{equation}
\mathsf{H}^{\mbox{elec}}~=~  \int^{\oplus}_{X}\mathsf{H}({\bf b},{\bf t}^{\rm e})_o~d{\bf b}.
\label{decomp}
\end{equation}
This result implies at once that the spectrum of $\mathsf{H}^{\mbox{elec}}$ is \emph{purely continuous} 
\begin{displaymath}
\sigma~=~\sigma(\mathsf{H}^{\mbox{elec}})~=~\bigcup_{\bf b} ~\sigma({\bf
  b})~\equiv [V_0,\infty)  
\end{displaymath}
where $V_0$ is the minimum value of $E({\bf b})_0$; in the diatomic
molecule case this is the minimum value of $E_0(r)$ (see Fig.
\ref{PEcurve}). $\mathsf{H}^{\mbox{elec}}$ has \emph{no} normalizable eigenvectors. We conclude therefore that the decomposition of the
molecular Hamiltonian into a nuclear kinetic energy operator
contribution, proportional to $\kappa^4$, and a remainder \emph{does not} yield molecular potential energy surfaces.

\section{Discussion}
\label{Disc}
The rigorous mathematical analysis of the original perturbation approach proposed by Born and Oppenheimer \cite{BO:27} for a molecular Hamiltonian with Coulombic interactions was initiated by Combes and co-workers \cite{CS:80,C:75,C:77} with results for the diatomic molecule. A perturbation expansion in powers of $\kappa$ leads to a singular perturbation problem because $\kappa$ is a coefficient of differential operators of the highest order in the problem; the resulting series expansion of the energy is an \emph{asymptotic} series, closely related to the WKB approximation. Some properties of the operator $\mathsf{H}^{\mbox{elec}}$, (equation \ref{decomp}), seem to have been first discussed in this work. Since the initial work of Combes, a considerable amount of mathematical work has been published using both time-independent and time-dependent techniques with developments for the polyatomic case; for a recent review of rigorous results about the separation of electronic and nuclear motions see Hagedorn and Joye \cite{HJ:07} which covers the literature to 2006. 

If it can be assumed that a) the electronic wavefunction
vanishes strongly outside a region close to a particular nuclear
geometry and b) that the electronic energy at the given geometry is an
isolated minimum, then it is possible to present a rigorous account of the separation of electronic and nuclear motion which
corresponds in some measure to the original Born-Oppenheimer
treatment;  such an account is provided for a polyatomic molecule by Klein and co-workers \cite{KMSW:92}. Because of the continuous spectrum of the electronic Hamiltonian $\mathsf{H}^{\mbox{elec}}$, it is not possible to use regular perturbation theory in the analysis; instead asymptotic expansion theory is used so that the result has essentially the character of a WKB approximation. Neither is it possible to consider rotational symmetry for the reasons outlined in \S \ref{modth}. However it is possible to impose inversion symmetry and the nature of the results derived under this requirement depend on the shape of the geometrical figure at the minimum energy configuration. If the geometry at the minimum is either linear or planar then inversion can be dealt with in terms of a single minimum in the electronic energy. If the geometry at the minimum is other than these two forms, inversion produces a second potential minimum and the problem must be dealt with as a two-minimum problem; then extra consideration is necessary to establish whether the two wells have negligible interaction so that only one of the wells need be considered for the nuclear motion. The nuclei are treated as distinguishable particles that can be numbered uniquely. The symmetry requirements on the total wavefunction that would arise from the invariance of the Hamiltonian operator under the permutation of identical nuclei are not
considered. 

It is similarly possible to consider such phenomena as Landau-Zener
crossing by using a time-dependent approach to the problem and looking at the relations between the electronic and nuclear parts of a wave packet \cite{HJ:99}. This is essentially a use of standard coherent state theory where again the nuclei are treated as distinguishable particles and the method is that of asymptotic expansion.

To summarize: what the rigorous work so far presented has done is to
show that if an electronic wavefunction has certain local
properties and if the nuclei can be treated as distinguishable
particles then the the eigenfunctions and eigenvalues of the
molecular (Coulomb) Hamiltonian can be obtained in a WKB expansion
in terms of $\kappa$ to arbitrary order.

In this paper we have attempted to discuss the Born approach to the
molecular theory in terms first set out by Combes \cite{C:75}. The
essential point is that the decomposition of the molecular Hamiltonian
(with center-of-mass contribution removed) into the nuclear kinetic
energy, proportional to $\kappa^4$ and a remainder, is specified by
equation (\ref{sepeqn}), not by (\ref{Hpert}), or in other words,
\emph{equation (\ref{Hpert}) cannot be written with an $=$
  sign}. Allowing the nuclear masses to increase without limit in
$\mathsf{H}^{\mbox{elec}}$ does not produce an operator with a
discrete spectrum since this would just cause the mass polarisation
term to vanish and the effective electronic mass to become the rest
mass. It is thus not possible to reduce the molecular Schr\"{o}dinger
equation to a system of coupled differential equations of classical
type for nuclei moving on 
potential energy surfaces, as suggested by Born, without a further approximation of an essentially empirical character. An \emph{extra choice}
of fixed nuclear positions must be made to give any discrete spectrum
and normalizable $L^2$ eigenfunctions.  In our view this choice, that
is, the introduction of the clamped-nuclei Hamiltonian, by hand, into
the molecular theory as in \S \ref{tradBO} is the essence of the ``Born-Oppenheimer approximation'' 
\begin{eqnarray}
\mathsf{H}^{\mbox{elec}}~&=&~  \int^{\oplus}_{X}\mathsf{H}({\bf b},{\bf t}^{\rm e})_o~d{\bf b}~\rightarrow~\mathsf{H}({\bf b},{\bf t}^{\rm e})_o\nonumber\\ &~&\mbox{Born-Oppenheimer approximation}. 
\label{BOapprox}
\end{eqnarray}
If the molecular Hamiltonian $\mathsf{H}$ were classical, the removal of the nuclear kinetic energy terms would indeed leave a Hamiltonian representing the electronic motion for stationary nuclei, as claimed by Born and Oppenheimer \cite{BO:27,fn1}. As we have seen, quantization of $\mathsf{H}$ changes the situation drastically, so an implicit appeal to the classical limit for the nuclei is required. The argument is a subtle one, for subsequently, once the classical energy surface has emerged, the nuclei are treated as quantum particles (though indistinguishability is rarely carried through); this can be seen from the complexity of the mathematical account given by Klein and co-workers \cite{KMSW:92}.

We have long argued that the solutions of the time-independent Schr\"{o}dinger equation for the molecular Hamiltonian are of limited interest for chemistry, being really only relevant to a quantum mechanical account of the physical properties (mainly spectroscopic) of atoms and diatomic molecules in the gas-phase. Towards the end of his life, P.-O. L\"{o}wdin made an extended study of a quantum mechanical definition of a molecule; in one of his late papers \cite{POL:89} he lamented
\begin{quote}
The Coulombic Hamiltonian $\mathsf{H}'$ does not provide much obvious information or guidance, since there is [\emph{sic}] no specific assignments of the electrons occurring in the systems to the atomic nuclei involved - hence there are no atoms, isomers, conformations etc. In particular one sees no \emph{molecular symmetry}, and one may even wonder where it comes from. Still it is evident that all this information must be contained somehow in the Coulombic Hamiltonian.
\end{quote}
In our view it is not at all evident that the Coulombic Hamiltonian \emph{on its own} will give rise to the chemically interesting features L\"{o}wdin required of it, nor will they be approachable by regular perturbation theory (supposedly convergent) starting from its eigenstates. Fundamental modifications of the quantum theory of the Coulomb Hamiltonian for a generic molecule have to be made for a chemically significant account of dipole moments, functional groups and isomerism, optical activity and so on 
\cite{W:76,WS:77,W:82,WS:03,SW:05b}. In other words one should not expect useful contact between the quantum theory of an \emph{isolated} molecule (which is what the eigenstates of the Coulombic Hamiltonian refer to) and a quantum account of \emph{individual} molecules, as met in ordinary chemical situations where \emph{persistent interactions} (due to the quantized electromagnetic field, other molecules in bulk media) and finite temperatures are the norm \cite{PG:83,W:88}.

The usual practice in computational chemistry in which clamped-nuclei
electronic energy calculations are used to define a potential energy
surface upon which a quantum mechanical nuclear motion problem is
solved, is a practice that is well defined mathematically. Solutions obtained to the equations formulated in this context do not
have the symmetry of the full internal motion Hamiltonian and their
relationship to solutions of that Hamiltonian is, as has been seen,
uncertain.

That said, the conventional account, treating formally identical
nuclei as identifiable particles when it seems chemically prudent to
do so, has enabled a coherent and progressive account of much chemical
experience to be provided. But it is not derived by continuous
approximations from the eigensolutions of the Schr\"{o}dinger equation for the molecular Coulomb Hamiltonian, requiring as it does an essential
empirical input. The tremendous success of the usual practice might
perhaps be best regarded as a tribute to the insight and ingenuity of
the practitioners for inventing an effective variant of quantum theory
for Chemistry.


\begin{thebibliography}{199}
\addcontentsline{toc}{section}{\numberline{}References}
\bibitem{EWK:44} H. Eyring, J. Walter and G.K. Kimball \emph{Quantum
  Chemistry}, John Wiley, London, (1944)
\bibitem {BO:27} M. Born and J.R. Oppenheimer, Ann. Phys. {\bf 84}, 457 (1927)
\bibitem{fn1} An English language translation of the original paper can be found at www.ulb.ac.be/cpm/people/scientists/bsutclif/main.html.
\bibitem{PW:35} L. Pauling and E. Bright Wilson \emph{Introduction to
  Quantum Mechanics}, McGraw-Hill, New York, (1935)
\bibitem{L:28}F. London in \emph{Probleme der modernen Physik}
  Ed. P. Debye, S. Hirsel, Leipzig, 104, (1928)  
\bibitem{B:51} M. Born, Nachr. Akad. Wiss. Goett. II. Math.-Phys. K1,
  {\bf 6}, 1 (1951) 
\bibitem{BH:54} M. Born and K. Huang, \emph{Dynamical Theory of Crystal Lattices}, Clarendon, Oxford, (1954)
\bibitem{CS:80}  J.-M. Combes and R. Seiler in
\emph{Quantum Dynamics of Molecules},  Ed. R. G. Woolley,  NATO ASI
B57, Plenum Press, New York,  435 (1980)
\bibitem{RS:78} M. Reed and B. Simon, B. \emph{Methods of Modern
Mathematical Physics, IV, Analysis of Operators}, Academic Press,
New York, (1978)
\bibitem{ZH:60} G.M. Zhislin, Tr. Mosk. Mat. Obs.  {\bf 9},
81 (1960); the full Russian text can be obtained from \url{http://mi.mathnet.ru/eng/mmo/v9/p81}.
\bibitem{Uch:67} J. Uchiyama, J. Pub. Res. Inst. Math. Sci. Kyoto
{\bf 2}, 117 (1966)
\bibitem{Th:81}  W. Thirring, W. \emph{A Course in Mathematical Physics, 3, Quantum Mechanics of Atoms and Molecules}, Tr. E.M. Harrell,
Springer-Verlag, New York (1981)
\bibitem{HYS:86} N.C. Handy, Y. Yamaguchi and H.F. Schaefer III, J. Chem. Phys. {\bf 84}, 4481 (1986)
\bibitem{HL:96} N.C.Handy and A.M. Lee, Chem. Phys. Letters, {\bf 252}, 425 (1996)
\bibitem{K:97} W. Kutzelnigg, Mol. Phys. {\bf 90}, 909 (1997)
\bibitem{AM:60} A. Messiah, \emph{Quantum Mechanics}, Vol. 2, North-Holland (1960)
\bibitem{MM:67} C.A. Mead and A. Moscowitz, Int. J. Quant. Chem. {\bf 1}, 243 (1967)
\bibitem{TOM:71} T. F. O'Malley,  Adv. At. Mol. Phys. {\bf 7}, 223 (1971)
\bibitem{HJM:87} H. J. Monkhorst, Phys. Rev. {\bf A36}, 1544 (1987)
\bibitem{LT:10} L. Lodi and J. Tennyson, J. Phys. B At. Mol. Opt. Phys. {\bf 43}, 133001 (2010) 
\bibitem{fn2} This is the analogue for operators with continuous spectra of the familiar result in linear algebra that the eigenspaces of an $n ~\times~ n$ matrix can be decomposed as a direct sum of a $n-$dimensional vector space. Alternatively, note that each of the spaces $H(P)$ transforms under a different representation of the translation group in 3-dimensions --- under translation by an amount $a$, an element of $H(P)$ is multiplied by the factor $e^{iPa}$. Thus the direct integral is the infinite-dimensional analogue of the decomposition of a finite-dimensional vector space on which a group acts as a direct sum of irreducible representations.
\bibitem{LMS:07} M. Loss, T. Miyao and H. Spohn, J. Funct. Anal. {\bf 243}, 353 (2007)
\bibitem{POL:88} P.-O. L\"{o}wdin, in \emph{Molecules in Physics, Chemistry and Biology}, Vol. 2, Ed. J. Maruani, Kluwer Academic Publishers, Dordrecht (1988)
\bibitem{SW:05a} B.T. Sutcliffe and R.G. Woolley, Phys. Chem. Chem. Phys. {\bf 7}, 3664 (2005)
\bibitem{FROL:99} A. M. Frolov, Phys. Rev. A., {\bf 59}, 4270 (1999)
\bibitem{K:51} T. Kato  Trans. Amer. Math. Soc. {\bf 70}, 212 (1951)
\bibitem{C:75} J.-M. Combes, International Symposium on Mathematical Physics (Kyoto University, Kyoto), Lecture Notes in Physics {\bf 39}, 467 (1975)
\bibitem{C:77} J.-M. Combes, Acta Phys. Austriaca Suppl., {\bf XVII}, 139 (1977)
\bibitem{HJ:07} G.A. Hagedorn and A. Joye in \emph{Spectral Theory and Mathematical Physics. A Festschrift in Honor of Barry Simon's 60th Birthday}, Eds. F. Gesztezy, P. Deift, C. Galvez, P. Perry and G.W. Schlag, Oxford University Press, London, 203 (2007)
\bibitem{KMSW:92} M. Klein, A. Martinez, R. Seiler, and X. P. Wang, 
Commun. Math. Phys., {\bf 143}, 607 (1992) 
\bibitem{HJ:99} G.A. Hagedorn and A. Joye, Revs. Math. Phys. {\bf 11}, 41 (1999)
\bibitem{POL:89} P.-O. L\"{o}wdin, Pure and Appl. Chem. {\bf 61}, 2065 (1989) 
\bibitem{W:76} R.G. Woolley, Adv. Phys. {\bf 25}, 27 (1976)
\bibitem{WS:77} R.G. Woolley and B.T. Sutcliffe, Chem. Phys. Letters, {\bf 45}, 393 (1977)
\bibitem{W:82} R.G. Woolley, Structure and Bonding, Springer-Verlag, Berlin, {\bf 52}, 1 (1982)
\bibitem{WS:03} R.G. Woolley and B.T. Sutcliffe, in \emph{Fundamental World of Quantum Chemistry}, Eds. E.J. Br\"{a}ndas and E.S. Kryachko, Kluwer Academic Publishers, Dordrecht, {\bf 1}, 21 (2003)
\bibitem{SW:05b} B.T. Sutcliffe and R.G. Woolley, Chem. Phys. Letters, {\bf 408}, 445 (2005)
\bibitem{PG:83} I. Prigogine and C. George, Proc. Natl. Acad. Sci. U.S.A. {\bf 80}, 4590 (1983)
\bibitem{W:88} R.G. Woolley in \emph{Molecules in Physics, Chemistry and Biology}, Vol. 1, Ed. J. Maruani, Kluwer Academic Publishers, Dordrecht (1988)
\end{thebibliography}
\end{document}